\documentclass[12pt]{article}
\pdfoutput=1
\usepackage{subfigure}
\usepackage{amssymb,amsmath}
\usepackage{graphicx}
\usepackage{color}
\usepackage[colorlinks=true
,urlcolor=blue
,citecolor=blue
,linkcolor=blue
,pagecolor=blue
,linktocpage=true
,pdfproducer=medialab
]{hyperref}
\usepackage[a4paper]{geometry}

\usepackage{placeins}

\makeatletter \renewcommand{\@dotsep}{10000} \makeatother

\def\mmess{M_{\rm mess}}
\mathchardef\mhyphen="2D

\newcommand{\beq}{\begin{equation}}
\newcommand{\eeq}{\end{equation}}
\newcommand{\bea}{\begin{eqnarray}}
\newcommand{\eea}{\end{eqnarray}}

\usepackage[nodisplayskipstretch]{setspace}

\begin{document}

\begin{titlepage}
\pagestyle{empty}

\begin{flushright}
FTPI-MINN-16/24
\end{flushright}

\vspace*{0.2in}
\begin{center}
{\Large \bf    Muon $g-2$ in GMSB with Adjoint Messengers
  }\\
\vspace{1cm}
{\bf  Ilia Gogoladze$^{a,}$\footnote{E-mail: ilia@bartol.udel.edu;\\
\hspace*{0.5cm} On leave of absence from Andronikashvili Institute
of Physics,  Tbilisi, Georgia.} and
Cem Salih $\ddot{\rm U}$n$^{b,}\hspace{0.05cm}$\footnote{E-mail: cemsalihun@uludag.edu.tr}}
\vspace{0.5cm}

{\it
$^a$Bartol Research Institute, Department of Physics and Astronomy,\\
University of Delaware, Newark, DE 19716, USA \\
$^b$Department of Physics, Uluda\~{g} University, TR16059 Bursa, Turkey
}

\end{center}

\vspace{0.5cm}
\begin{abstract}
\noindent
We explored the sparticle mass spectrum in light of the muon $g-2$ anomaly  and the little hierarchy problem in a class of gauge mediated supersymmetry  breaking model. Here   the messenger fields transform in the adjoint representation of the Standard Model gauge symmetry. To avoid unacceptably light right-handed slepton masses  the standard model is supplemented by additional $U(1)_{B-L}$ gauge  symmetry.  Considering a non-zero $U(1)_{B-L}$ D-term leads to an additional contribution to the soft supersymmetry breaking mass terms which makes the right-handed slepton masses compatible with the current experimental bounds.
We show that in the framework of  $\Lambda_{3}<0$  and  $\mu < 0$,   the muon $g-2$ anomaly and  the observed 125 GeV Higgs boson mass a can be simultaneously accommodated. The slepton masses in this case are predicted to lie in the few hundred GeV range, which can be tested at LHC.  Despite the heavy colored  spectrum  the the little hierarchy problem in this model can be ameliorated and electroweak fine tuning parameter can be as low as 10 or so.
\end{abstract}

\end{titlepage}



\section{Introduction}
\label{ch:introduction}

Even though minimal supersymmetric Standard Model (MSSM) can still be fit consistently with the current experimental results, the  discovery  of  the  Higgs  boson  with  a  mass $ \sim 125$ GeV \cite{:2012gk,:2012gu} and absence of  signal  of the supersymmetric particles   \cite{atlas11} at LHC bring severe constraints on the sparticle mass spectrum.  It require rather heavy sparticles. Such a heavy mass spectrum for the sparticles  leads to   the question of   the little hierarchy problem \cite{b5}.  In the models with gauge mediated supersymmetry breaking (GMSB) \cite{Dine:1993yw,  Meade:2008wd} the little hierarchy problem becomes more severe compare to the the gravity mediation supersymmetry    breaking scenario  \cite{Chamseddine:1982jx}. In general, the trilinear SSB A-terms in GMSB scenarios are relatively small  at the messenger scale,
even if an additional sector is added to generate the $\mu/B\mu$ terms~\cite{Komargodski:2008ax}.
Because of the small A-terms, accommodating the light CP-even Higgs boson mass around 125~GeV requires a stop mass
in the  multi-TeV range~\cite{Ajaib:2012vc}.  While a large trilinear SSB  terms  allows relatively light stop quark solutions in gravity mediation supersymmetry    breaking scenario  to be consistent with the Higgs boson mass constraint \cite{Demir:2014jqa}.
On the other hand, a multi-TeV top squark has a very strong influence
on the sparticle spectrum ~\cite{Ajaib:2012vc},  if we assume that the messenger fields reside in the SU(5) representations
such as  $5+\bar5$ or $10+\overline{10}$. This case is called minimal GMSB scenario, since it  is the simplest
scenario that preserves gauge coupling unification of the  minimal supersymmetric standard model (MSSM)
and  provides non-zero SSB mass terms for all supersymmetric particles.
It is often assumed that all  messenger fields have a universal mass in this simplest model.
Even if one assumes a large mass splitting among the colored and non-colored messenger fields,
the sparticle mass spectrum cannot be entirely separated, since all  fields  from $5+\bar5$
(or $10+\overline{10})$  representation have non-zero hypercharge, and they can generate non-zero masses
through hypercharge interactions. This means that in these models the maximal splitting among sfermion SSB mass terms
cannot exceed the ratio  of corresponding fermion hypercharges.
Therefore, the colored and non colored sparticle mass spectra are closely linked here.
For instance, if we have a multi-TeV mass  top squark in the minimal GMSB scenario,
then the whole SUSY sparticle spectrum is also around the TeV scale~\cite{Ajaib:2012vc}.
Note that  $t$-$b$-$\tau$ Yukawa coupling unification can be realized in these models and it provides
a specific spectrum for sparticle masses~\cite{Gogoladze:2015tfa} which tested at future experiments.

It was shown a while ago that a GMSB model could be realized  having the messenger fields  in the adjoint representations of $SU(3)_{C}$  and  $SU(2)_{L}$ \cite{Han:1998pa}, we call this scenario as GMSB-Adj. The messenger fields in these representations do not carry any hypercharge, and hence there is no common SSB terms generated for the colored and non-colored sectors. Thus, one can have a light mass spectrum for the left handed sleptons, while the stops and othe colored particles still can  have multi-TeV  masses. Indeed, this model suffers from inconsistently light right handed  sleptons.  Since in this scenario messenger fields does not have hypercharge as the result  the right-handed sleptons are generated at high loop level  at the $\mmess$ scale and is negligible compare othes sfermion masses.  As we know slepton mass can increase through RGE evaluation. But   there is no enough contribution from the RGE flaw  to realize sleptons heavier than about 100 GeV at the low scale. To overcome this problem, one can consider non-zero D-term contributions \cite{Gogoladze:2016grr} which generate SSB mass term for the right-handed sleptons at $\mmess$. A detailed analyses have been performed in a recent study \cite{Gogoladze:2016grr}, and it has been shown that a consistent slepton spectrum can be realized when a non-zero $U(1)_{B-L}$ D-term contribution is consider around  the messenger scale.  In this set up, the stops are still heavier than about 2 TeV, while the all slepton  are  rather light, 100 GeV or so.

Besides the right handed  sleptons in the  GMSB-Adj scenario  bino also does not obtain SSB mass term at one loop level.
But RGE evaluation make bino mass in the  $O$ (GeV)$-100$(GeV) interval.  On the other hand  there is not sever constraint for bino mass from the experiment.  For instance   existence  of  a  near  massless  bino,  however,  would  contribute  to $\Delta N_{eff}=N_{eff}-N_{eff, SM}$,  The reason for this is that the essentially massless bino
decouples from the thermal background around the same time as the neutrinos.  The
decoupling temperature also depends on the slepton mass which we take around the
weak scale.  However, if the slepton mass increases, the decoupling temperature also
increases, e.g., if the slepton mass is 10 TeV, then the decoupling temperature will be
O(GeV). The  BICEP2 data~\cite{Ade:2014xna}  requires a larger $\Delta N_{\rm eff}$(=0.81 $\pm 0.25$) in order
to reconcile with the Planck data~\cite{Dvorkin:2014lea}.  Future data hopefully will settle this issue.

  It is interesting to note that  GMSB-Adj scenario \cite{Gogoladze:2016grr} there are parameters space where the little hierarchy problem is  ameliorated. Particularly it was found in ref.  \cite{Gogoladze:2016grr}  that  electroweak fine-tuning measure $\Delta_{EW}$   can be as low as $\Delta_{EW} \gtrsim 50$.  Detailed discussion about $\Delta_{EW}$ see in  \cite{Baer:2012mv} and references therein.

A light slepton spectrum is also favored, since it yields significant contributions to the muon anomalous magnetic moment (here after muon $g-2$). However, as shown in Ref. \cite{Gogoladze:2016grr}, it is not possible to resolve the muon $g-2$ problem in GMSB-Adj when it was consider $\Lambda_{3}>0$  and  $\mu > 0$. There reason is that starting with tine mass for bino at the messenger   scale the RGE evaluation   leads to negative value for  bino mass  at the electroweak scale.
 On the other hand the contributions from the sparticles to muon $g-2$ is proportional to the combination of $\mu M_{i}$, where $i=1,2$ represent masses of the gauginos of $U(1)_{Y}$ and $SU(2)_{L}$ respectively. A light Bino along with light sleptons significantly contribute to muon $g-2$, but having bino in the model with negative sign \cite{Gogoladze:2016grr} turns such contributions to be destructive in muon $g-2$ calculation. Even though $M_{2}$ provides contributions enhancing muon $g-2$, they are not enough to suppress those from the bino loop \cite{Gogoladze:2016grr}.

In this paper we continue investigation of model   presented in Ref. \cite{Gogoladze:2016grr}. Here we consider the parameter space when $\Lambda_{3}<0$  and  $\mu < 0$.  It leads to have at low scale $M_i<0$ and  $\mu < 0$ while $\mu M_{i}>0$. So,  as we will show in this case there is no negative contribution to the muon  $g-2$ from bino and/or wino loop.  We also consider electro week fine-tuning for this scenario.
  The outline of the paper is as follows: We briefly discuss the essential features of the model and the situation of muon $g-2$ in Section \ref{sec:model}. After we summarize our scanning procedure and the experimental constraints imposed in our analyses in Secton \ref{sec:scan}, we present our results in Section \ref{sec:results}. Finally we summarize and conclude our results in Section \ref{sec:conc}.

\section{ The  GMSB-Adj model}
\label{sec:model}

We summarize the essential features of GMSB-Adj in this section. A detailed description is given in Refs. \cite{Han:1998pa}. In GMSB models, SUSY is broken in a hidden sector with a singlet field $S$ through the following superpotential

\begin{equation}
 W \supset (m_3+\lambda_3\, S) {\rm Tr}(\Sigma_3^2)  + (m_8+\lambda_8\, S) {\rm Tr} (\Sigma_8^2).
\end{equation}
where $m_{3}$ and $m_{8}$ are the messenger scales relevant to triplet and octet messengers denoted with $\Sigma_{3}$ and $\Sigma_{8}$ respectively. For simplicity it can be assume that $m_{3}=m_{8}=\mmess$. The $S$ field breaks SUSY when its $F_{S}$ component develops a non-zero vacuum expectation value (VEV) denoted as $\langle F_{S} \rangle$. This VEV generates a mass splitting between bosonic and fermionic components of the messenger superfields as follows:

\begin{equation}
 m_{b_{i}}=\mmess \sqrt{1\pm \dfrac{\Lambda_{i}}{\mmess}}~,\hspace{0.3cm} m_{f_{i}}=\mmess
 \label{messengermass}
\end{equation}
where $m_{b_{i}}$ and $m_{f_{i}}$ with $i=3,8$ represent the bosonic and the fermionic components of $\Sigma
_{3}$ and $\Sigma_{8}$ respectively. The mass difference between the bosons and fermions are expressed in terms of $\Lambda_{i}$, where $\Lambda_{3}=\lambda_{3}\langle F_{S}\rangle/\mmess$ and $\Lambda_{8}=\lambda_{8}\langle F_{S}\rangle/\mmess$. The messenger fields $\Sigma_{3}$ and $\Sigma_{8}$ decouple at $\mmess$ and generate SSB mass terms. These mass terms are generated at one-loop for the gauginos as

\beq
M_{1}=0 ~,\hspace{0.3cm}M_{2}\simeq \frac{g_{2}^{2}}{16\pi^{2}}2\Lambda_{3}~\hspace{0.3cm}
M_{3}\simeq \frac{g_{3}^{2}}{16\pi^{2}}3\Lambda_{8}~,
\label{BCgaugino}
\eeq
where $M_{i}$ with $i=1,2,3$ stand for gauginos associated with $U(1)_{Y}$, $SU(2)_{L}$, and $SU(3)_{C}$ respectively. $\Lambda_{3}=\lambda_{3}\langle F_{S}\rangle/\mmess$, and similarly $\Lambda_{8}=\lambda_{8}\langle F_{S}\rangle/\mmess$. Although the bino mass vanishes at the messenger scale,
it is generated below the messenger scale through the RGE evolution ~\cite{Yamada:1993ga}. The
dominant contribution to the bino mass has the following form:

\beq
 (16\pi^2)^2\frac{d}{dt}M_{1}= 2 g_1^2\left( \frac{27}{5}g_2^2M_2+ \frac{88}{5}g_3M_3\right) + \frac{26}{5}y_{t}A_{t}.
 \label{bino}
\eeq

The scalar SSB masses are generated at two loops and for GMSB-Adj  case we have  \cite{Han:1998pa}

\begin{eqnarray}
&& m^{2}_{\tilde{Q}} \simeq \dfrac{2}{(16\pi^{2})^{2}}\left[ \dfrac{4}{3}g_{3}^{4}3\Lambda^{2}_{8}+\dfrac{3}{4}g_{2}^{4}2\Lambda_{3}^{2} \right] \nonumber \\
&& m^{2}_{\tilde{U}} =  m^{2}_{\tilde{D}} \simeq  \dfrac{2}{(16\pi^{2})^{2}}\left[\dfrac{4}{3}g_{3}^{4}3\Lambda_{8}^{2} \right] \nonumber \\
&& m^{2}_{L} \simeq \dfrac{2}{(16\pi^{2})^{2}}\left[ \dfrac{3}{4}g_{2}^{4}2\Lambda_{3}^{2} \right] \nonumber \\
&& m^{2}_{H_{u}} =  m_{H_{d}}^{2} = m^{2}_{L} \nonumber \\
&& m_{E}^{2} =  0~.
\label{BCscalar}
\end{eqnarray}
where we have used the standard notation for the MSSM fields.

The right-handed slepton masses will be generated at a higher loop level, and thus, they vanish at the messenger scale.
However, they are generated below the messenger scale from the RGE evolution.
 On the other hand, experimental constraints  require that the sleptons must be heavier than 100~GeV or so. In order to generate right-handed slepton masses of order 100~GeV or heavier   through RGE flaw,
some of the other sparticles should be around 100~TeV or so, which makes supersymmetry much less motivated for solving the gauge hierarchy problem.

To avoid this problem it was proposed to   consider an extension of  the SM gauge symmetry with $U(1)_{B-L}$, which is one of the most natural extension of the SM gauge symmetry. We assume that the $U(1)_{B-L}$ symmetry is spontaneously broken not far below the messenger scale.  Thus, below the messenger scale the RGE evolution is   the  the one associated with the MSSM. It is very natural to consider non-zero D-term contribution \cite{Drees:1986vd} to the MSSM scalar SSB mass term. In this case we have
\begin{equation}
m_{\phi}^{2}=(m_{\phi}^{2})_{{\rm GMSB-Adj}}+(e_{B-L}D)^{2}
\end{equation}
where $\phi$ represents the scalars and $(m_{\phi}^{2})_{{\rm GMSB-Adj}}$ is their masses generated by the messenger scales as given in Eq.(\ref{BCscalar}). $D$ stands for non-zero D-term contributions, and $e_{B-L}$ denotes the $B-L$ charge of the  field.

\section{Scanning Procedure and Experimental Constraints}
\label{sec:scan}

For our scan over the fundamental parameter space of GMSB with the adjoint messengers, we employed ISAJET 7.84 package~\cite{Paige:2003mg}
supplied with appropriate boundary conditions at $M_{\rm Mess}$. In this package, the weak-scale values of gauge and Yukawa
couplings are evolved from $M_Z$ to $M_{\mathrm{Mess}}$ via the MSSM RGEs in the $\overline{DR}$ regularization scheme.
For simplicity, we do not include the Dirac neutrino Yukawa coupling in the RGEs, whose contribution is expected to be small.

The SSB terms are induced at the messenger scale and we set them according to
Eqs.~(\ref{BCgaugino}) and (\ref{BCscalar}).
From $M_{\mathrm{Mess}}$ the SSB parameters, along with the gauge and Yukawa couplings, are evolved down to the weak scale $M_Z$.
In the evolution of Yukawa couplings the SUSY threshold
corrections~\cite{Pierce:1996zz} are taken into account at the common scale $M_{{\rm SUSY}} = \sqrt{m_{\tilde{t}_{L}}m_{\tilde{t}_{R}}}$,
where $m_{\tilde{t}_{L}}$ and $m_{\tilde{t}_{R}}$ are the soft masses of the third generation left and right-handed top squarks respectively.

We have performed random scans over the model parameters in the following range:
\bea
\label{parameterRange}
-10^{7}\leq & \Lambda_{3} & \leq 10^{3} {\rm GeV} \nonumber \\
10^{3}\leq & \Lambda_{8} & \leq 10^{7} {\rm GeV} \nonumber \\
10^{3}\leq & M_{{\rm Mess}} & \leq 10^{16} {\rm GeV} \\
0 \leq & D & \leq 2000 {\rm GeV} \nonumber \\
2 \leq & \tan\beta & \leq 60 \nonumber\\
 \mu < 0  & {\rm and }&c_{\rm grav}=1~.  \nonumber
\eea

Regarding the MSSM parameter $\mu$, its magnitude but not the sign is determined by the radiative electroweak symmetry breaking (REWSB).
In our model we set ${\rm sign}(\mu)=-1$. Finally, we employ the current central value for the top mass, $m_{t}=173.3$~GeV.
Our results are not too sensitive to one or two sigma variation of $m_{t}$~\cite{Gogoladze:2011db}.

It is also well known that weak-scale SUSY can accommodate the $2-3 \, \sigma$ discrepancy between the measurement of muon  $g-2$  by the BNL~\cite{Bennett:2006fi}
experiment and its value predicted by the SM. It requires the existence of relatively light smuon and gaugino (wino or bino) \cite{Moroi:1995yh}. BNL has measured an excess of $\sim 3.6 \, \sigma \, (2.4 \, \sigma)$ in muon $g-2$, using $e^+e^- \rightarrow$ hadrons (hadronically decaying $\tau$) data. Various theoretical computations within the SM~ \cite{Hagiwara:2011af} have been performed by different groups to explain this excess, but to no avail. The deviation in $g-2$ from the SM prediction is:
\begin{eqnarray}
\Delta a_{\mu}\equiv a_{\mu}({\rm exp})-a_{\mu}({\rm SM})= (28.6 \pm 8.0) \times 10^{-10}\,.
\label{g-2}
\end{eqnarray}
In this paper we are looking to the   GMSB-Adj parameter space that resolves the $g-2$ anomaly.

In scanning the parameter space, we employ the Metropolis-Hastings
algorithm as described in Ref.~\cite{Belanger:2009ti}. The data points collected all satisfy
the requirement of radiative electroweak symmetry breaking (REWSB). We successively apply mass bounds
including the Higgs boson~\cite{:2012gu,:2012gk} and gluino masses~\cite{gluinoLHC},
and the constraints from the rare decay processes $B_s \rightarrow \mu^+ \mu^-$~\cite{BsMuMu},
$b \rightarrow s \gamma$~\cite{Amhis:2012bh} and $B_u\rightarrow\tau \nu_{\tau}$~\cite{Asner:2010qj}.
The constraints are summarized below in Table~\ref{table1}.
\begin{table}[h]\centering
\begin{tabular}{rlc}
$   123\, {\rm~GeV} \leq m_h \leq127$ \,{\rm~GeV} &
\\
$ m_{\tilde{g}} \geq 1.8$ \,{\rm TeV} & \\
$0.8\times 10^{-9} \leq{\rm BR}(B_s \rightarrow \mu^+ \mu^-)
  \leq 6.2 \times10^{-9} \;(2\sigma)$ &
\\
$2.99 \times 10^{-4} \leq
  {\rm BR}(b \rightarrow s \gamma)
  \leq 3.87 \times 10^{-4} \; (2\sigma)$ &
\\
$0.15 \leq \frac{
 {\rm BR}(B_u\rightarrow\tau \nu_{\tau})_{\rm MSSM}}
 {{\rm BR}(B_u\rightarrow \tau \nu_{\tau})_{\rm SM}}
        \leq 2.41 \; (3\sigma)$ &
\end{tabular}
\caption{Phenomenological constraints implemented in our study.}
\label{table1}
\end{table}

\section{Results}
\label{sec:results}

\begin{figure}[ht!]
\centering
\subfigure{\includegraphics[scale=0.8]{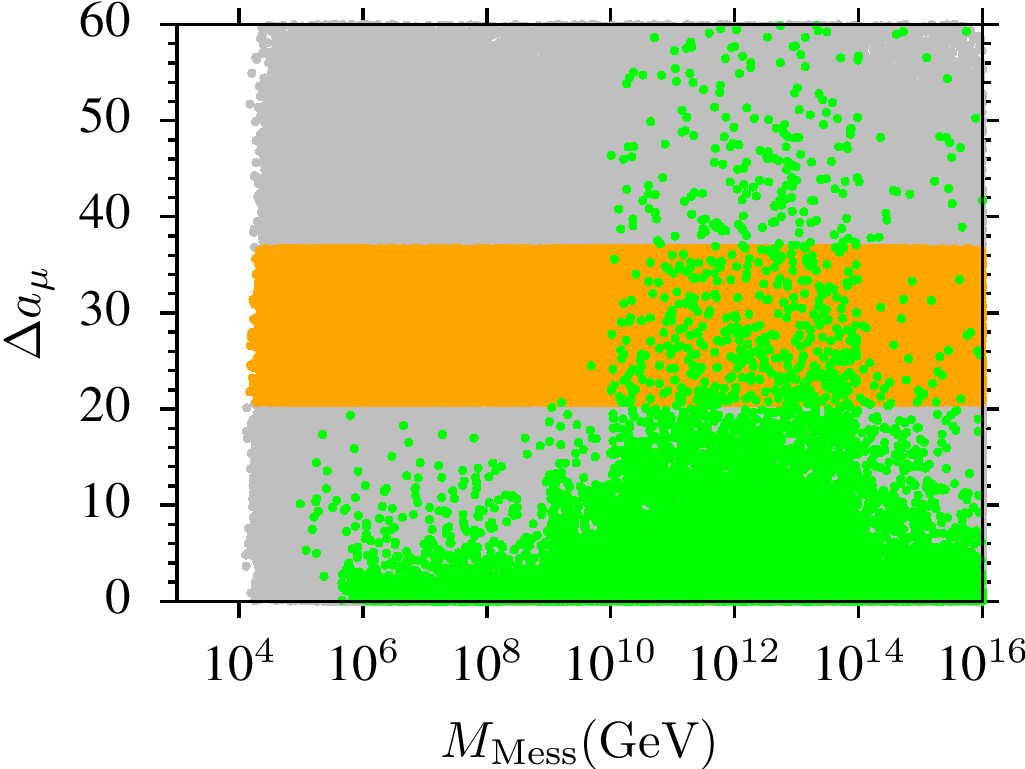}}
\subfigure{\includegraphics[scale=0.8]{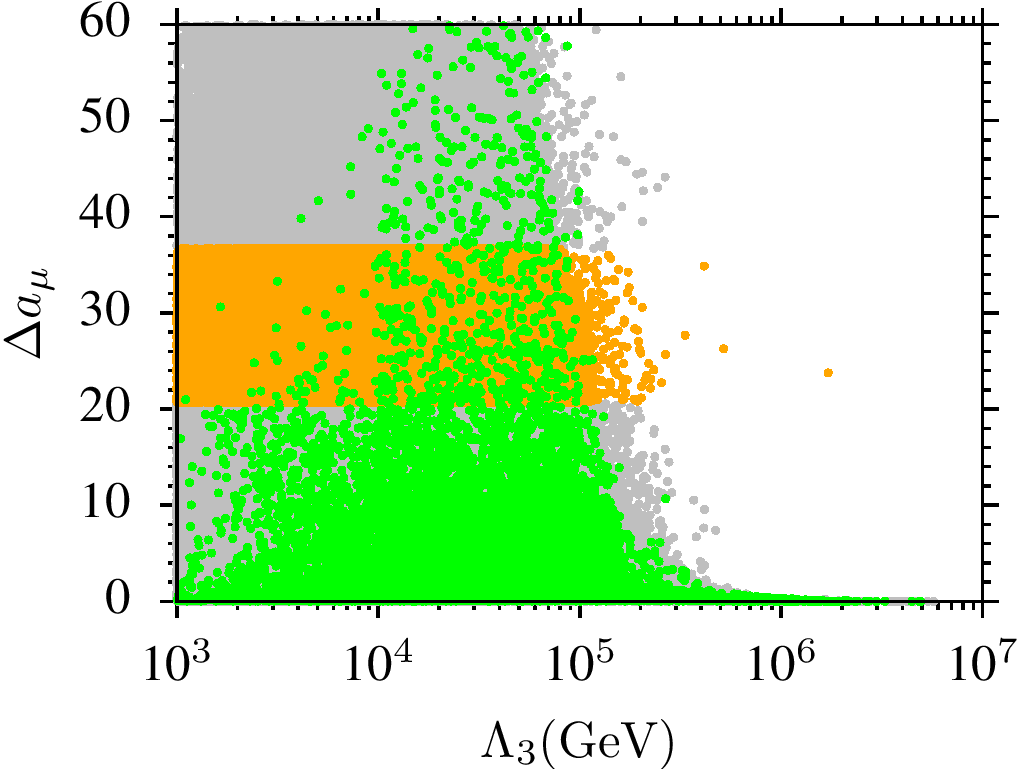}}
\subfigure{\includegraphics[scale=0.8]{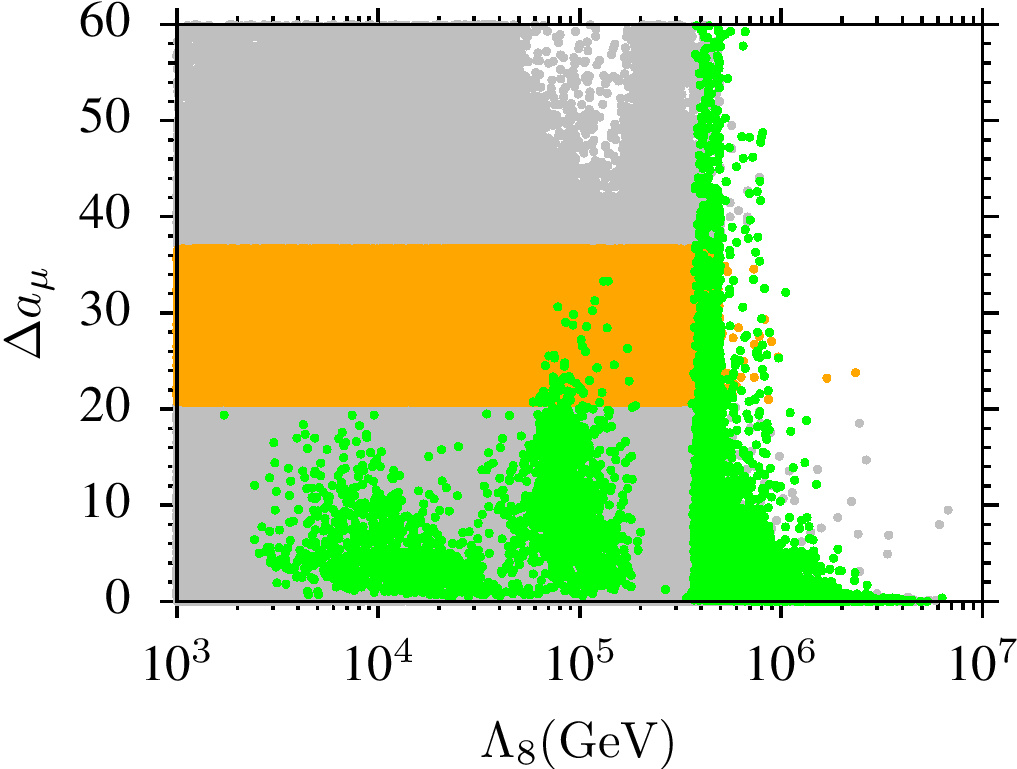}}
\subfigure{\includegraphics[scale=0.8]{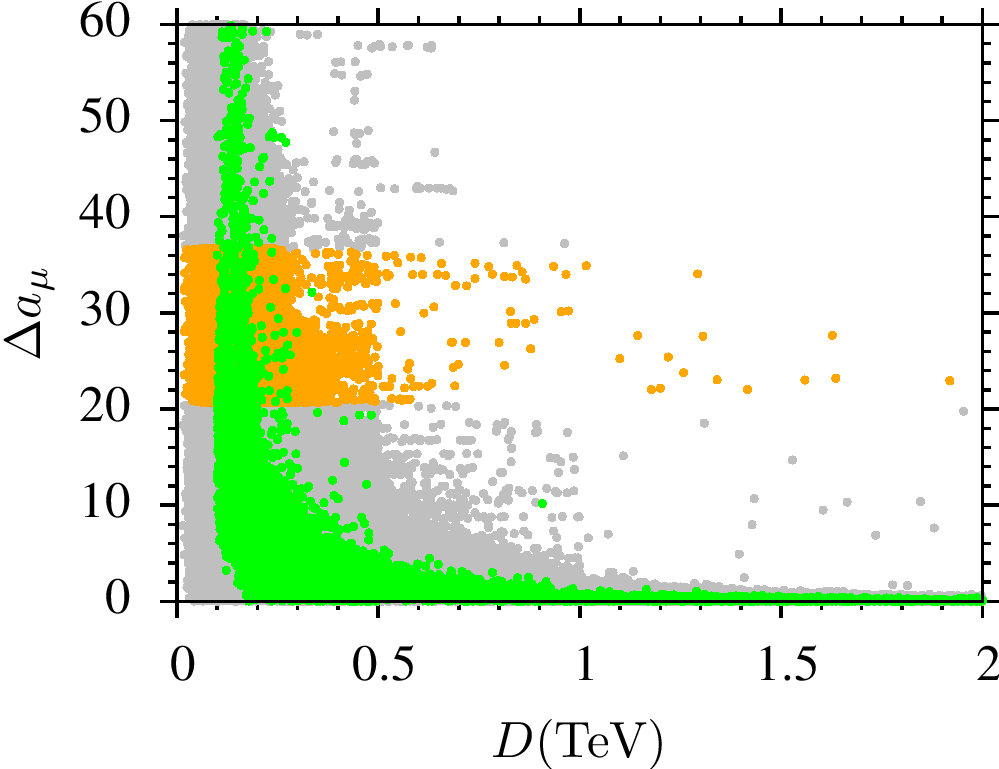}}
\subfigure{\includegraphics[scale=0.8]{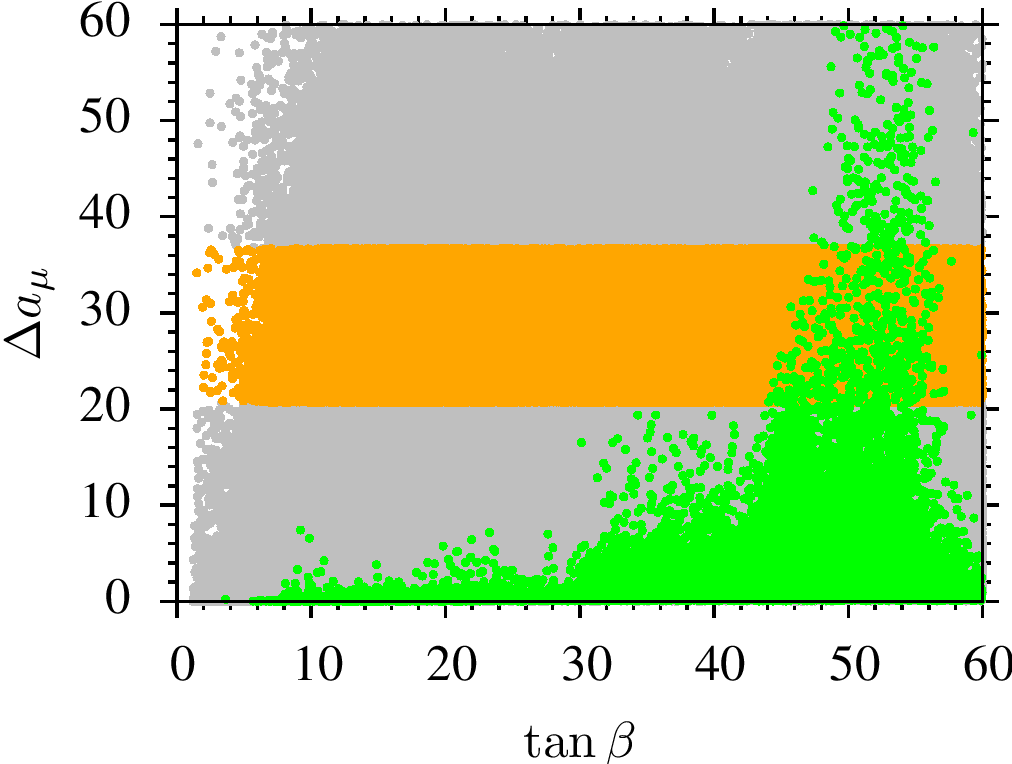}}
\subfigure{\includegraphics[scale=0.8]{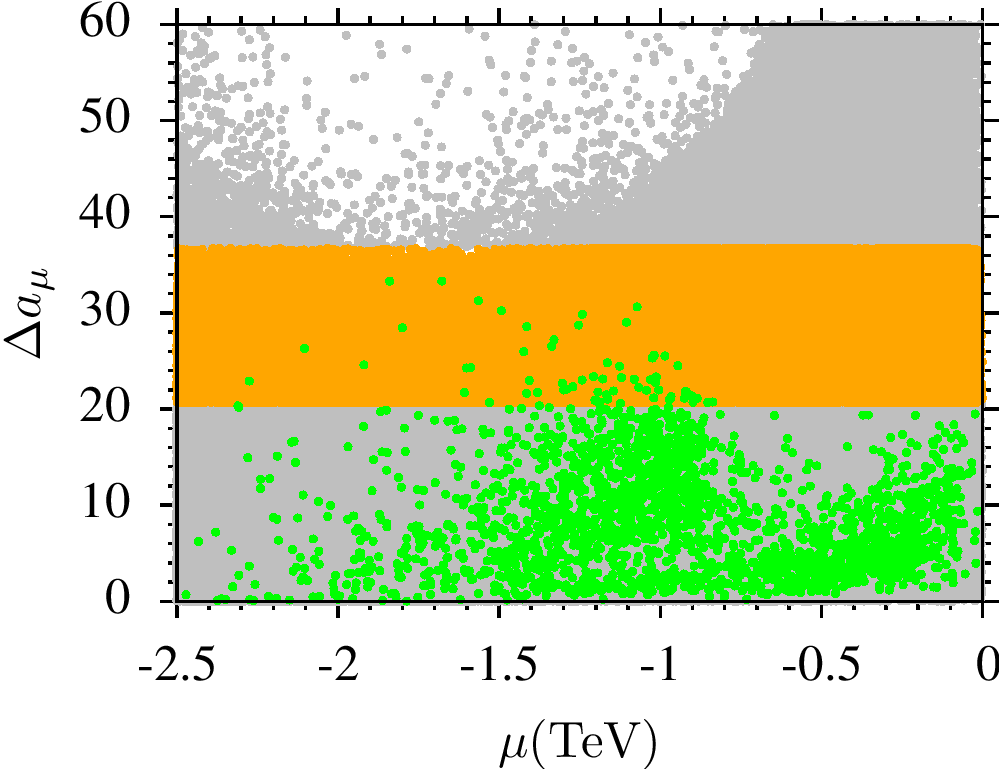}}
\caption{Plots in $\Delta a_{\mu}-\mmess$, $\Delta a_{\mu}-\Lambda_{3}$, $\Delta a_{\mu}-\Lambda_{8}$, $\Delta a_{\mu}-D$, $\Delta a_{\mu}-\tan\beta$, and $\Delta a_{\mu}-\mu$. All points are consistent with REWSB. Green points are consistent with the experimental constraints including the Higgs boson mass. The yellow band is an independent subset and it represents regions where $\Delta a_{\mu}$ would bring theory and experiment to within $1\sigma$.}
\label{fig1}
\end{figure}

\begin{figure}[ht!]
\centering
\subfigure{\includegraphics[scale=0.8]{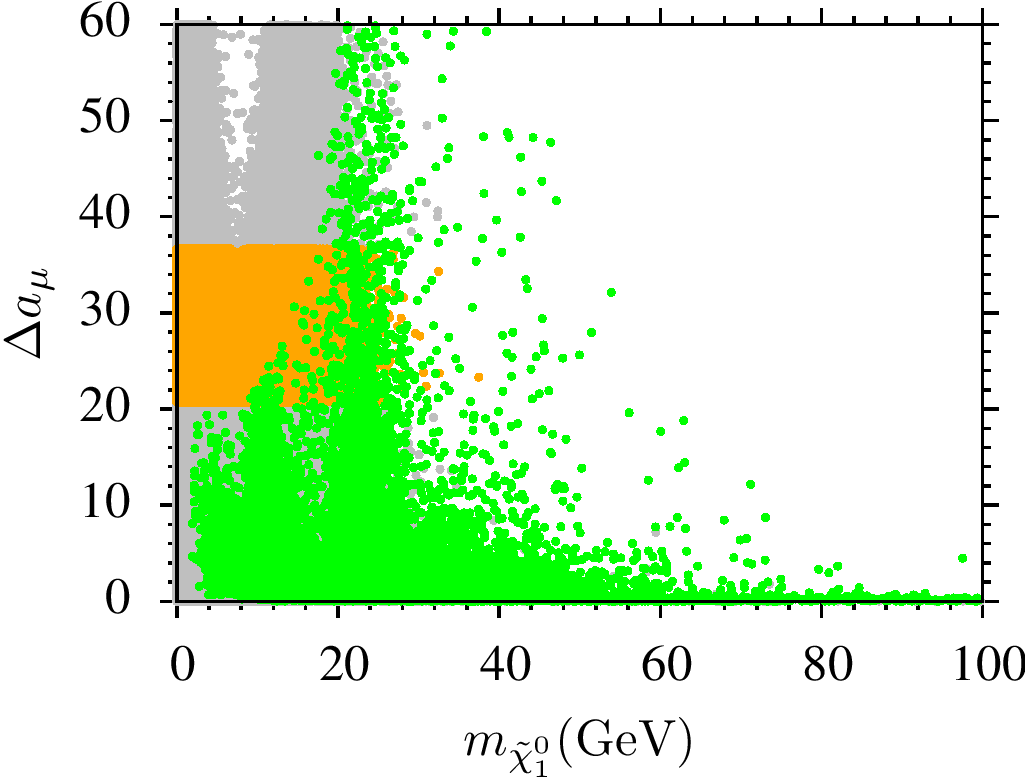}}
\subfigure{\includegraphics[scale=0.8]{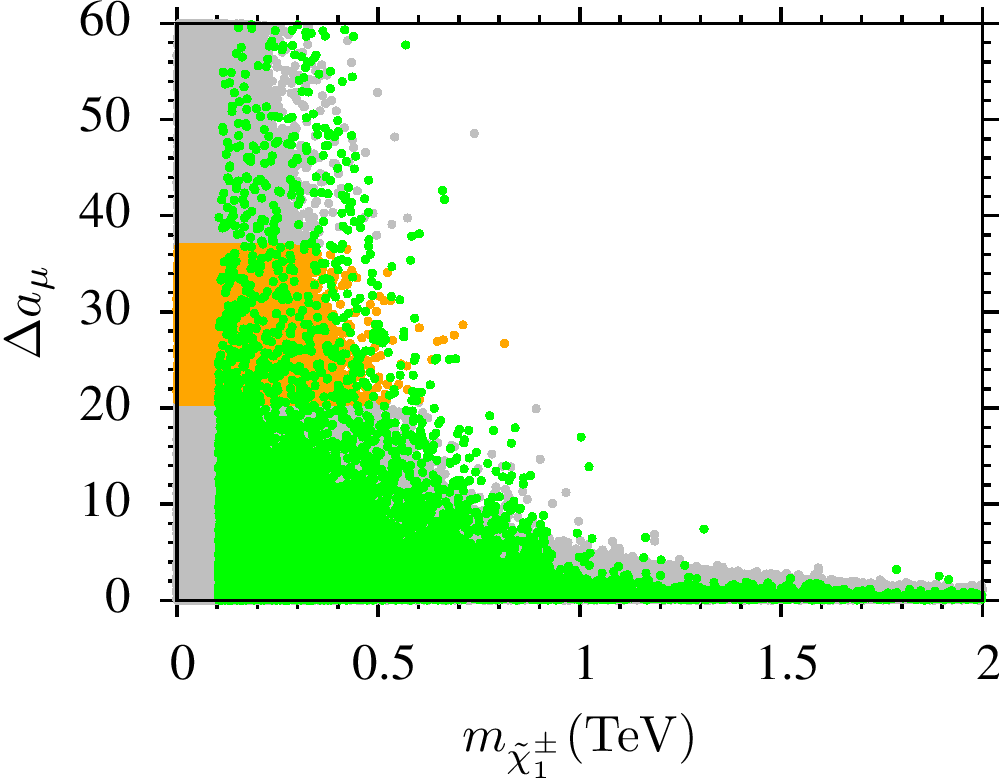}}
\subfigure{\includegraphics[scale=0.8]{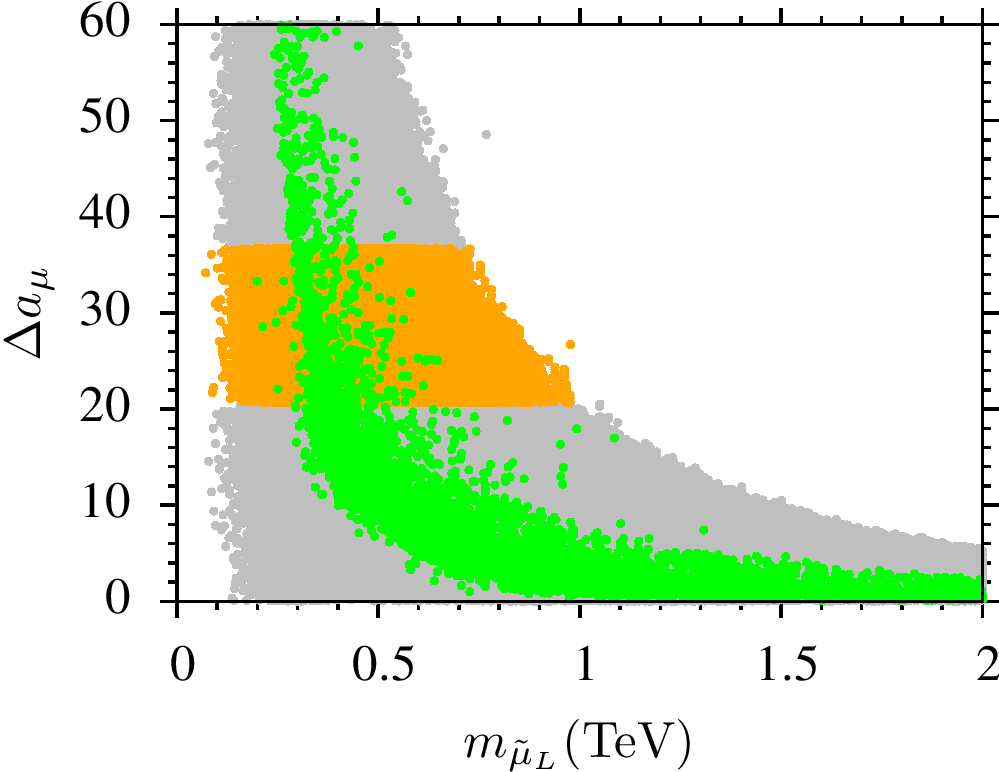}}
\subfigure{\includegraphics[scale=0.8]{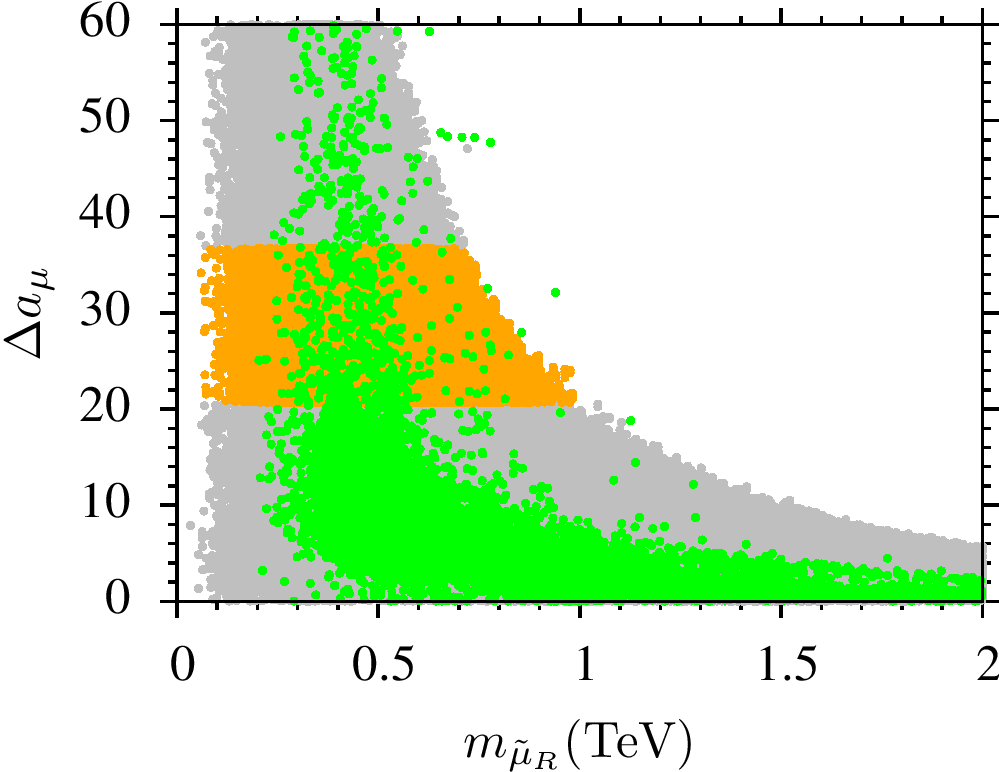}}
\caption{Plots in $\Delta a_{\mu}-m_{\tilde{\chi}_{1}^{0}}$, $\Delta a_{\mu}-m_{\tilde{\chi}_{1}^{\pm}}$, $\Delta a_{\mu}-m_{\tilde{\mu}_{L}}$, and $\Delta a_{\mu}-m_{\tilde{\mu}_{R}}$. The color coding is the same as Figure \ref{fig1}. }
\label{fig2}
\end{figure}

\begin{figure}[h!]
\centering
\subfigure{\includegraphics[scale=0.8]{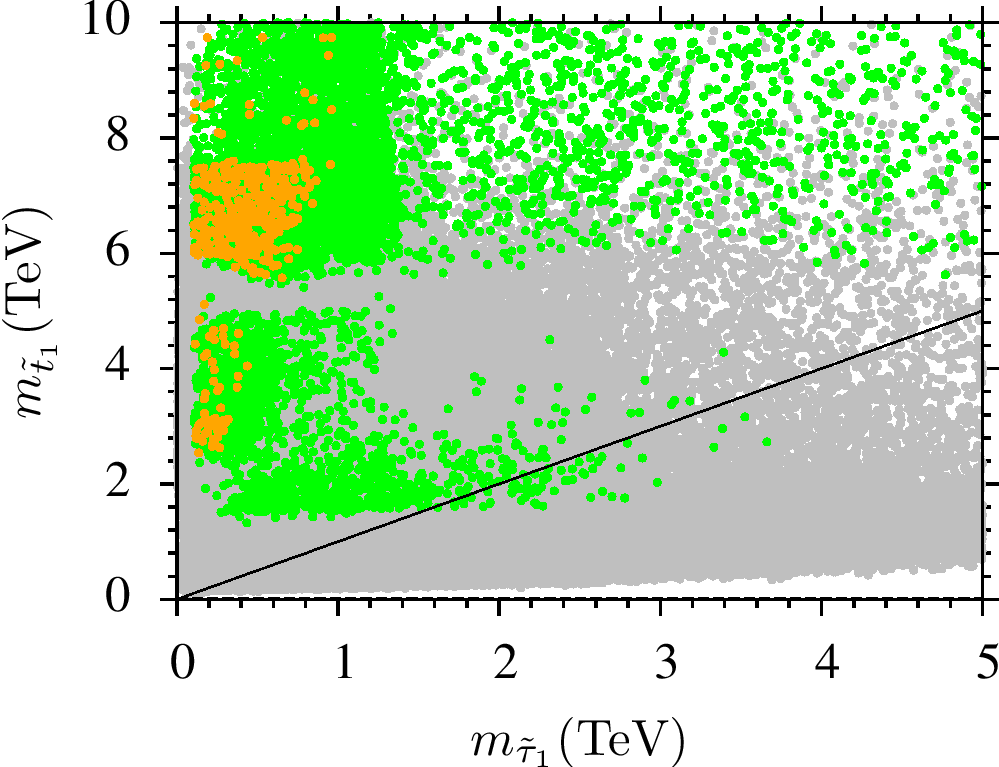}}
\subfigure{\includegraphics[scale=0.8]{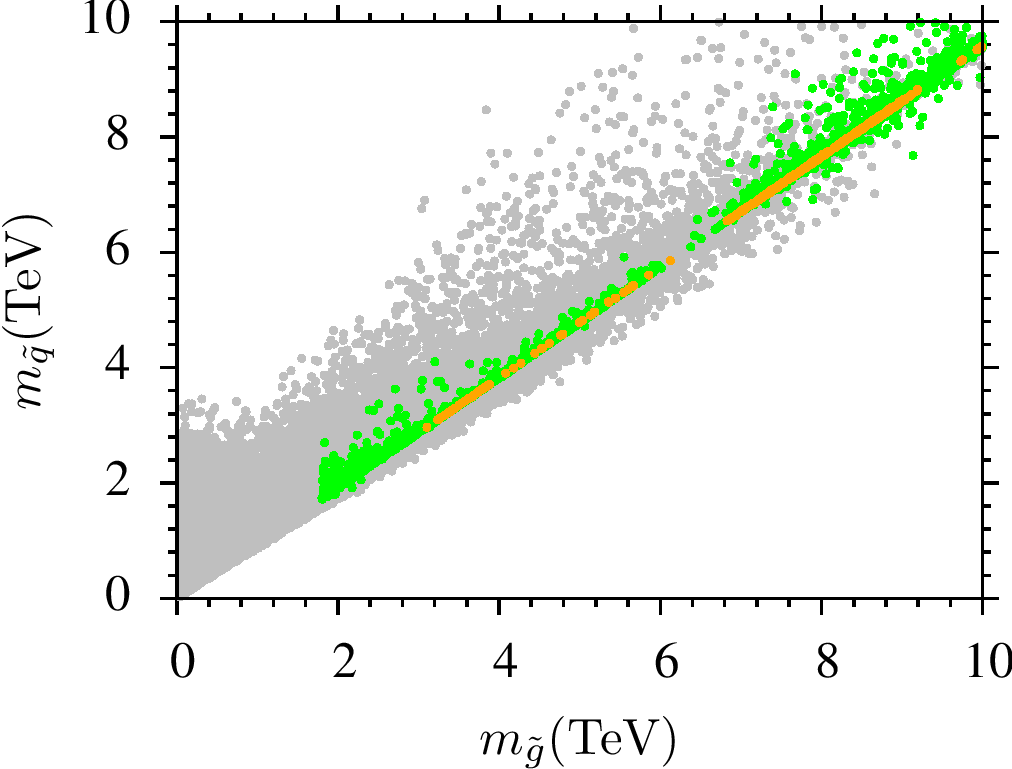}}
\subfigure{\includegraphics[scale=0.8]{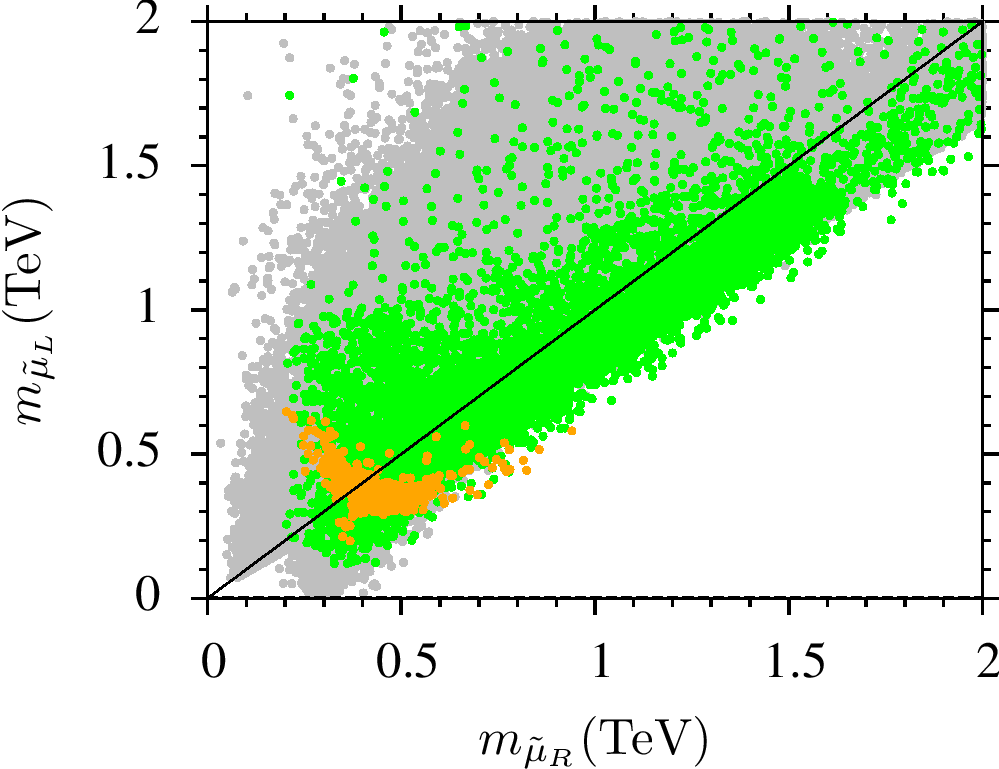}}
\subfigure{\includegraphics[scale=0.8]{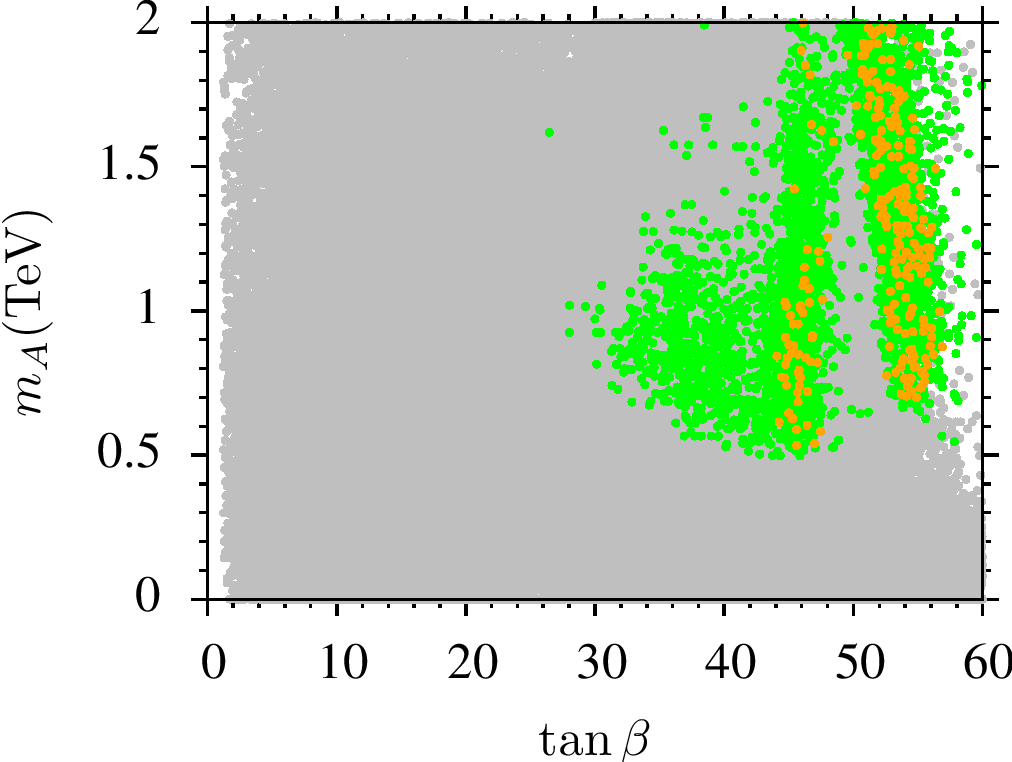}}
\caption{Plots in $m_{\tilde{t}_{1}}-m_{\tilde{\tau}_{1}}$, $m_{\tilde{q}}-m_{\tilde{g}}$, $m_{\tilde{\mu}_{L}}-m_{\tilde{\mu}_{R}}$, and $m_{A}-\tan\beta$. The color coding is the same as Figure \ref{fig1} except that yellow points are a subset of green and they represent values of $\Delta a_{\mu}$ that would bring theory and experiment to within $1\sigma$.}
\label{fig3}
\end{figure}

\begin{figure}[t!]
\centering
\includegraphics[scale=1]{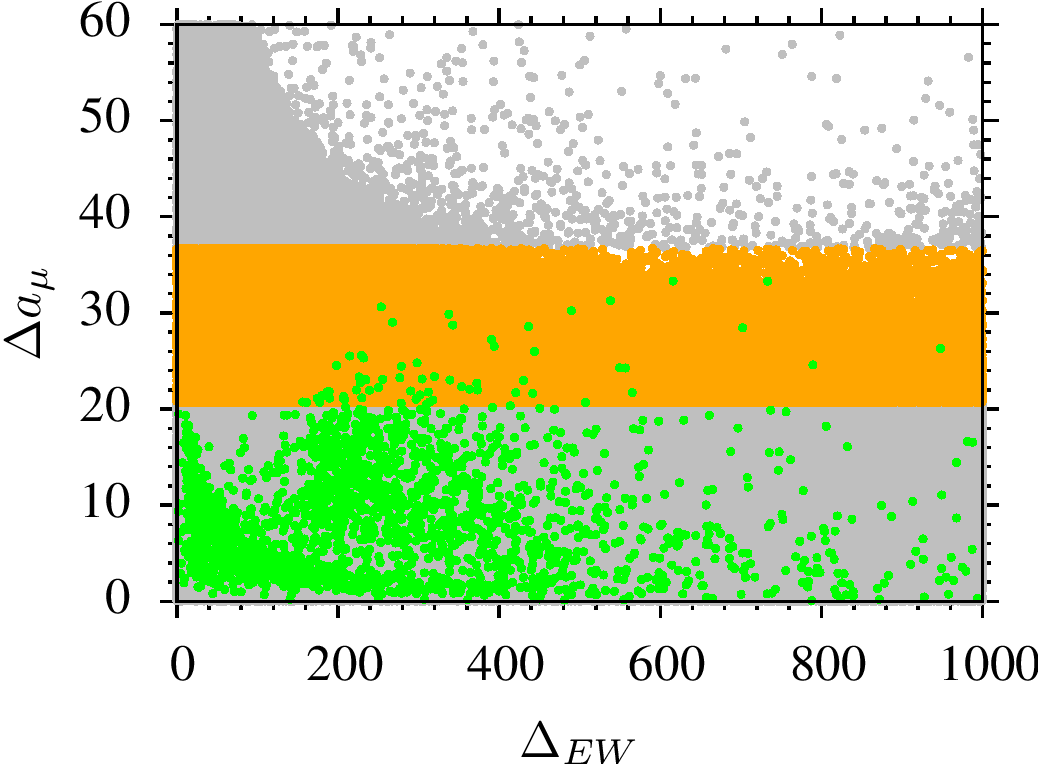}
\caption{Plot in the $\Delta a_{\mu}-\Delta_{EW}$ plane. The color coding is the same as Figure \ref{fig1}.}
\label{fig4}
\end{figure}

We present our results from the scan we performed over the fundamental parameter space listed in the previous section. We first focus on the regions, which are allowed by the experimental constraints mentioned in the previous section and can solve muon $g-2$ anomaly.  Figure \ref{fig1} represents our results with the plots in  $\Delta a_{\mu}-\mmess$, $\Delta a_{\mu}-\Lambda_{3}$, $\Delta a_{\mu}-\Lambda_{8}$, $\Delta a_{\mu}-D$, $\Delta a_{\mu}-\tan\beta$, and $\Delta a_{\mu}-\mu$. All points are consistent with REWSB. Green points are consistent with the experimental constraints including the Higgs boson mass. The yellow points satisfy the muon
 $g-2$ constraint given in Eq. (\ref{g-2}). The  $\Delta a_{\mu}-\mmess$, $\mmess$ plane shows that $\mmess$ needs to be larger than about $10^{10}$ GeV in order to have muon $g-2$ within $1\sigma$ interval from experimental observation.  We found that muon $g-2$ constraint put upper bound for $\Lambda_3<10^5$ GeV, see $\Delta a_{\mu}-\Lambda_{3}$ plane. Note that we plotted the absolute value of $\Lambda_{3}$, which is set to be negative in our scan. On the other hand, $\Lambda_{8}$  need to be more then $10^5$ GeV in order to accommodate light CP even higgs bound and  muon $g-2$ anomaly. As we mentioned
above  in order to have experimentally acceptable right handed slepton mass we need to have non zero $U(1)_{B-L}$ D-term. It need to be more then 100 GeV and when $D>500$ GeV the supersymmetric contribution to muon $g-2$ anomaly become negligible.  The $\Delta a_{\mu}-\tan\beta$ plane shows that one needs  to have $45<\tan\beta <55$ in order to satisfy constraint given in Section \ref{sec:scan}. In  the $\Delta a_{\mu}-\mu$ plane we are shoing solution with relatively small value for $|\mu|<2.5$. The reason is that we are seeking  for the solution when little hierarchy problem is not so severe.  As we can see from Figure 1 that there are many solution for $\mu \sim 100$  GeV or so,  which means that in the GMSB-Adj scenario we do not have strong fine tuning constraints for electroweak symmetry breaking. We will discuss more about fine tuning and its relation to muon $g-2$ when we present  $\Delta a_{\mu}-\Delta_{EW}$ plot.

Figure \ref{fig2} displays our results for the sparticle mass spectrum relevant to the SUSY contributions to muon $g-2$ with plots in $\Delta a_{\mu}-m_{\tilde{\chi}_{1}^{0}}$, $\Delta a_{\mu}-m_{\tilde{\chi}_{1}^{\pm}}$, $\Delta a_{\mu}-m_{\tilde{\mu}_{L}}$, and $\Delta a_{\mu}-m_{\tilde{\mu}_{R}}$. The color coding is the same as Figure \ref{fig1}. The lightest neutralino is not heavier than about 200 GeV, and the muon $g-2$ requirements bounds its mass further at about 50 GeV from below as seen in the $\Delta a_{\mu}-m_{\tilde{\chi}_{1}^{0}}$ plane. A similar result for the chargino mass is shown in $\Delta a_{\mu}-m_{\tilde{\chi}_{1}^{\pm}}$. Its mass is required to be lighter than about 800 GeV \cite{Ajaib:2015yma} to have significant contributions to muon $g-2$. The bottom panels of Figure \ref{fig2} displays the mass spectrum for both left and right-handed smuons. Since they can have non-zero masses at $\mmess$ because of non-zero D-term contributions, they can be as heavy as 2 TeV, and muon $g-2$ implies $m_{\tilde{\mu}_{R}} \lesssim 700$ GeV and $m_{\tilde{\mu}_{L}}\lesssim 900$ GeV.

We continue to present mass spectrum results in Figure \ref{fig3} with plots in $m_{\tilde{t}_{1}}-m_{\tilde{\tau}_{1}}$, $m_{\tilde{q}}-m_{\tilde{g}}$, $m_{\tilde{\mu}_{L}}-m_{\tilde{\mu}_{R}}$, and $m_{A}-\tan\beta$. The color coding is the same as Figure \ref{fig1} except that yellow points are a subset of green and they represent values of $\Delta a_{\mu}$ that would bring muon $g-2$  theoretical and experimental result   within $1\sigma$ uncertainty. The Higgs boson mass itself requires $m_{\tilde{t}}\gtrsim 2$ TeV (green), and muon $g-2$ lifts the bound on stops to about 3 TeV (yellow) as seen from the $m_{\tilde{t}_{1}}-m_{\tilde{\tau}_{1}}$ plane. In contrast to stops, staus can be as light as about 100 GeV in the same region. The other colored supersymmetric particles are shown in the $m_{\tilde{q}}-m_{\tilde{g}}$, and $m_{\tilde{q}}, m_{\tilde{g}} \gtrsim 4$ TeV. We present the smuon masses once more in the $m_{\tilde{\mu}_{L}}-m_{\tilde{\mu}_{R}}$ in comparison to each other. The D-term contribution allows the right-handed smuon to be heavier than the left-handed smuon for some solutions (below the diagonal line). Finally, we show the possible mass range for the CP-odd Higgs boson versus $\tan\beta$. The exclusion limit on CP-odd mass for large $\tan\beta$ can be expressed as $m_{A}\gtrsim 800$ GeV \cite{Khachatryan:2014wca}. Even after the exclusion, there are still plenty of solutions with heavy $A-$boson, which are hard to be detected at the experiments.

Finally, we discuss the fine-tuning in our model for the cases in which the muon $g-2$ discrepancy is resolved in Figure \ref{fig4} with the plot in the $\Delta a_{\mu}-\Delta_{EW}$ plane. The color coding is the same as Figure \ref{fig1}. For  $\Delta_{EW}$ = electroweak scale fine tuning parameter calculation we used
latest (7.84) version of  ISAJET~\cite{Paige:2003mg}. This calculation includes  the one-loop effective potential contributions to the tree level MSSM Higgs potential, the Z boson mass is given by the  relation:
\beq
\frac{M_Z^2}{2} =
\frac{(m_{H_d}^2+\Sigma_d^d)-(m_{H_u}^2+\Sigma_u^u)\tan^2\beta}{\tan^2\beta
-1} -\mu^2 \; .
\label{eq:mssmmu}
\eeq
The $\Sigma$'s stand for the contributions arising from the one-loop effective potential (for more details see ref.~\cite{Baer:2012mv}).
All parameters  in Eq.~(\ref{eq:mssmmu}) are defined at the weak scale $M_{EW}$.

In order to measure the EW scale fine-tuning condition associated with the little hierarchy problem, the following definitions are used~\cite{Baer:2012mv}:
\beq
 C_{H_d}\equiv |m_{H_d}^2/(\tan^2\beta -1)|,\,\, C_{H_u}\equiv
|-m_{H_u}^2\tan^2\beta /(\tan^2\beta -1)|, \, \, C_\mu\equiv |-\mu^2 |,
\label{cc1}
\eeq
 with each $C_{\Sigma_{u,d}^{u,d} (i)}$  less than some characteristic value of order $M_Z^2$.
Here, $i$ labels the SM and supersymmetric particles that contribute to the one-loop Higgs potential.
For the fine-tuning condition we have
\beq
 \Delta_{\rm EW}\equiv {\rm max}(C_i )/(M_Z^2/2).
\label{eq:ewft}
\eeq
Note that Eq.~(\ref{eq:ewft}) defines the fine-tuning  condition at $M_{EW}$ without addressing
the question of the origin of the parameters that are involved.

The $\Delta a_{\mu}-\Delta_{EW}$ plane shows that there is possible to have  $\Delta_{EW}$ 50 or so. This means that in GMSB-Adj scenario little hierarchy problem can be ameliorated significantly. it was shown  in Ref. \cite{Gogoladze:2013wva}
 that the little hierarchy problem  can be largely resolved if the ratio between $SU(2)_L$ and $SU(3)_c$
gaugino masses satisfy the asymptotic relation $M_2/M_3\approx 3$ at the GUT scale, which corresponds to $\Lambda_3/\Lambda_8>2$ ratio in GMSB-Adj scenario. We can see from Figure 1 that there are solutions satisfying this condition.
In this case the leading contributions to  $m^{2}_{H_{u}}$ through  RGE  evolution, which are proportional to $M_2$ and $M_3$, can cancel each other. This allows for large  values of  $M_2$ and $M_3$ in
 our scenario while keeping the value of $m^{2}_{H_{u}}$ relatively small. On the other hand, large values of $M_2$ and $M_3$ yield a heavy stop quark ($>$ few TeV) which is necessary in order to accommodate   $m_h \simeq 125$ GeV. Since in our scenario bino mass does not depend on $\Lambda_3$ and $\Lambda_8$ and it is mostly generated through RGE evaluation, it can be around 20 GeV or so, (see Figure 2). Thus, bino loop can be the dominant contributor in muon $g-2$ calculation. As we can see, requiring to have significant contributions to muon $g-2$  makes $\Delta_{EW}>200$, which we consider moderate and quite acceptable under the fine-tuning condition.

Note that in our scenario, Gravitino can be either lightest supersymmetric particle (LSP) or next to lightest supersymmetric  particle (NLSP). For a detailed study on these possibilities see in ref. \cite{Allahverdi:2014bva}. We will not discuss much about gravitino as the dark matter candidate  since there is not much difference here compared to the other GMSB scenario. We only remark that in GMSB-Adj scenario LSP gravitino can be $(O)30$ eV, or $(O)$ keV or so. Also in our scenario    one could invoke axions as cold dark matter and have exactly the same supersymmetric spectrum.

\begin{table}[ht!]
\centering
\scalebox{1.0}{
\begin{tabular}{|l|ccc|}
\hline
                 & Point 1 & Point 2 & Point 3 \\
\hline
\hline
$\Lambda_{3}$  & $0.68\times 10^{4}$  & $0.17\times 10^{4}$ & $0.20\times 10^{4}$  \\
$\Lambda_{8}$  & $0.43\times 10^{4}$  & $0.73\times 10^{5}$ & $0.47\times 10^{4}$ \\
$M_{\rm mess}$ & $0.29\times 10^{10}$ & $0.83\times 10^{15}$ & $0.12\times 10^{7}$ \\
$\tan\beta$    & 35 & 45 & 36\\
$D$            & 143 & 100 & 138\\
\hline
$\mu$          & -89 & -1028 & -69 \\
$\Delta_{EW}$  & 10 & 233  &\textbf{10}\\
$\Delta a_{\mu}$ &$20\times 10^{-10}$ & $26\times 10^{-10}$ & $20\times 10^{-10}$\\
\hline
$m_h$           & 126 & 125 & 126\\
$m_H$           & 783  & 810  & 678 \\
$m_A$           & 778 & 805  & 673\\
$m_{H^{\pm}}$   & 788 & 817 & 685 \\

\hline
$m_{\tilde{\chi}^0_{1,2}}$
                 & 5, 157  & \textbf{10}, 799 & 3, 15  \\

$m_{\tilde{\chi}^0_{3,4}}$
                 & 997, 999  & 2046, 2047 & 853, 854 \\

$m_{\tilde{\chi}^{\pm}_{1,2}}$
                & 158, 1000 & 80, 2047 & 16, 856 \\

$m_{\tilde{g}}$  & 1844 & 3434 & 1879 \\
\hline $m_{ \tilde{u}_{L,R}}$
                 & 1758,1792  & 3278, 3303 & 1774, 1800   \\
$m_{\tilde{t}_{1,2}}$
                 & 1574, 1657 & 2823, 2920 & 1648, 1700 \\
\hline $m_{ \tilde{d}_{L,R}}$
                 & 1760, 1758  & 3279, 3290 & 1776, 1778  \\
$m_{\tilde{b}_{1,2}}$
                 & 1606, 1689  & 2839, 2928 & 1651, 1731 \\
\hline
$m_{\tilde{\nu}_{e,\mu}}$
                 & 397 & 319 & 284 \\
$m_{\tilde{\nu}_{\tau}}$
                 & 394 & 408  & 286 \\
\hline
$m_{ \tilde{\mu}_{L,R}}$
                & 406, 412  & 331, 398 & 297, 312  \\
$m_{\tilde{\tau}_{1,2}}$
                & 312, 467  & 176, 591 & 187, 375 \\
\hline
$m_{\tilde{G}}$  & $5\times 10^{-6}$  & 146 & $1.36\times 10^{-6}$\\
\hline
\end{tabular}}
\caption{Benchmark points for exemplifying our results. All masses are in GeV. All points are chosen as to be consistent with the experimental constraints given in Section \ref{sec:scan}. Point 1 exemplifies a solution with gravitino LSP, while Point 2 depicts one with neutralino LSP. Point 3 displays a solution with a keV scale gravitino LSP with $\Delta_{EW} \sim 10$ and $\Delta a_{\mu} \sim 20\times 10^{-10}$.}
\label{benchsgmsb}
\end{table}

Before concluding our results, we provide a table for three benchmark points exemplifying our results. All masses are in GeV. All points are chosen as to be consistent with the experimental constraints given in Section \ref{sec:scan}. Point 1 exemplifies a solution with a keV scale gravitino LSP with $\Delta_{EW} \sim 10$, while Point 2 depicts one with neutralino LSP. Both points yield light sleptons ($m_{\tilde{\tau}_{1}}\sim 193, 176$ GeV). Even though the GMSB models typically yield gravitino LSP, GMSB-Adj predicts also neutralino LSP solutions when $\mmess$ is large. Point 2 illustrates such a solution with $\Delta_{EW}\sim 200$ despite $\mmess \sim 10^{15}$ GeV. Point 3 displays a solution which needs, like Point 1, very low fine-tuning ($\Delta_{EW}\sim 10$) at the electroweak scale. In this case, the gravitino LSP is of mass at the order eV. Even though the muon $g-2$ results of Points 1 and 3 are slightly out of the $1\sigma$ band, it is still comparable with the experimental results since $\Delta a_{\mu} \sim 20\times 10^{-10}$. Recall that we wrote the absolute value of $\Lambda_{3}$ in Table \ref{benchsgmsb}, which is set to be negative in our scans.

\section{Conclusion}
\label{sec:conc}

We explored the sparticle mass spectrum in light of the muon $g-2$ and the little hierarchy problem in a class of gauge mediated SUSY breaking models in which the messenger fields are resided in the adjoint representation of $SU(3)_{C}\times SU(2)_{L}$. To avoid unacceptably light right-handed sleptons we introduce a non-zero $U(1)_{B-L}$ D-term. In this framework, the SSB mass terms for the colored and non-colored sectors are generated different independent parameters, and these two sectors are completely untied, since the hypercharge interactions are absent. These models are also favored by the low fine-tuning requirement in resolution of the little hierarchy problem. We found that cancelations between the terms proportional to $\Lambda_{8}$ and $\Lambda_{3}$ can yield low fine-tuning at any value of $\mmess$.

In addition, a negative $\Lambda_{3}$ along with a negative $\mu-$term allows significant SUSY contributions which accommodate the muon $g-2$ resolution with the 125 GeV Higgs boson mass. The solutions for the muon $g-2$ resolution restrict the fundamental parameters as $\mmess \gtrsim 10^{8}$ GeV, and $\Lambda_{3}\lesssim 10^{5}$ GeV, while $\tan\beta \gtrsim 35$. The $D-$term contributions are strictly bounded from above as $D \lesssim 300$ GeV, and muon $g-2$ results sharply drop to about zero for values beyond this bound.  In addition the neutralino mass should be lighter than about 30 GeV, while the chargino can be as heavy as about 600 GeV. Smuons also cannot be heavier than about 600 GeV. On the other hand, the color sector is rather heavy as $m_{\tilde{t}} \gtrsim 2$ TeV and $m_{\tilde{g}}\gtrsim 4$ TeV. The CP-odd higgs mass can be realized in a wide range from about 500 GeV to a few TeV. Despite the heavy spectrum in the colored sector, we have  solutions  $\Delta_{EW} \approx 10$ satisfying all experimental constraints and makes muon $g-2$ within $2\sigma$ deviation from current experimental bound.

\vspace{-0.4cm}
\section*{Acknowledgments}
This work is supported in part by  Bartol Research Institute (I.G.), and The Scientific and Technological Research Council of Turkey (TUBITAK) Grant no. MFAG-114F461 (CS\"{U}). Part of the numerical calculations reported in this paper
were performed at the National Academic Network and Information Center (ULAKBIM) of TUBITAK, High Performance and Grid Computing Center (TRUBA Resources).

\end{document}